\documentclass[aps,prb,twocolumn,superscriptaddress,showpacs]{revtex4}
\usepackage{graphicx}
\usepackage{epsfig}
\usepackage{dcolumn}
\usepackage{amsmath,amsfonts,amssymb,amsthm,verbatim}
\usepackage[mathlines]{lineno}% Enable numbering of text and display math
\usepackage{subfigure}
\def\be{\begin{equation}}
\def\ee{\end{equation}}
\def\ba{\begin{eqnarray}}
\def\ea{\end{eqnarray}}

\def\eg{{\it e.g.}}

\begin{document}
\title{Chiral $p$-wave superconductivity in Sb(111) thin films close to Van Hove singularities}
\author{Jin-Qin Huang}
\affiliation{State Key Laboratory of Optoelectronic Materials and Technologies,	School of Physics, Sun Yat-sen University, Guangzhou 510275, China}
\author{Chuang-Han Hsu}
\affiliation{Centre for Advanced 2D Materials and Graphene Research Centre,	National University of Singapore, 6 Science Drive 2, Singapore 117546}
\affiliation{Department of Physics, National University of Singapore, 2 Science Drive 3, Singapore 117542}
\author{Hsin Lin}
\affiliation{Centre for Advanced 2D Materials and Graphene Research Centre,	National University of Singapore, 6 Science Drive 2, Singapore 117546}
\affiliation{Department of Physics, National University of Singapore, 2 Science Drive 3, Singapore 117542}
\author{Dao-Xin Yao}
\email{yaodaox@mail.sysu.edu.cn}
\affiliation{State Key Laboratory of Optoelectronic Materials and Technologies,	School of Physics, Sun Yat-sen University, Guangzhou 510275, China}
\author{Wei-Feng Tsai}
\email{wftsai@mail.nsysu.edu.tw}
\affiliation{Department of Physics, National Sun Yat-sen University, Kaohsiung 80424, Taiwan}
\affiliation{Centre for Advanced 2D Materials and Graphene Research Centre,	National University of Singapore, 6 Science Drive 2, Singapore 117546}
\affiliation{Department of Physics, National University of Singapore, 2 Science Drive 3, Singapore 117542}
\date{\today}
\begin{abstract}
We theoretically investigate the development of unconventional superconductivity in the Sb(111) thin film when its Fermi level is tuned to near type-II Van Hove singularities (VHS), which locate at non-time-reversal invariant momenta. Via patch renormalization group analysis, we show that the leading instability is a chiral $p+ip$-wave superconducting order. The origin of such pairing relies on the hexagonal structure	of the VHS and strong spin-orbit coupling, resulting in the anisotropy of the electron-electron scattering to provide an attractive channel. Our study hence suggests that superconducting Sb thin films originated from VHS physics may host Majorana zero modes in the magnetic vortices and provides another application perspective to such material. 
\end{abstract}
\maketitle

\section{Introduction}
Since the discovery of the time-reversal symmetry (TRS) protected, Z$_2$ topological insulators (TIs), searching for more realizable topological states of matter becomes one of the attractive tasks in condensed matter community.\cite{Hasan10,Qi11,Moore10,Ando13} Using K-theory or Altland-Zirnbauer scheme, people even propose a ``periodic table'' for classifying topological insulators and superconductors (SCs).\cite{Kitaev09,Schnyder08} One class of the earlier known topological superconductors is the so-called 2D chiral (TRS-broken) superconductors. Chiral SCs can exhibit several intriguing properties due to their non-trivial topology of band structures, such as gapless chiral edge modes that carry quantized thermal current\cite{Volovik97,Senthil99} and Majorana zero modes bound in the vortices.\cite{Kopnin91,Read00,Ivanov01,Tewari07} Although there are proposed candidate chiral SCs like Sr$_2$RuO$_4$, the experimental evidence is still not definitive.\cite{Mackenzie03} More potential materials are therefore needed.

There are basically two approaches to achieve 2D chiral superconductivity: 1) such unconventional SC is triggered internally by electron-electron interactions\cite{Laughlin94,Jiang08,bschaffer07,Baskaran10,Lin11,Thomale14,Kallin15}; 2) such order is induced externally at the interface by the proximity effect of a conventional SC to a quantum (anomalous) Hall insulator or Rashba system.\cite{Qi10,Alicea12} In the first approach, ``strong'' correlation usually plays an essential role. Thus, a promising way which caught people's eye is to consider the physics around Van Hove singuralities (VHS). At 2D VHS, the density of states (DOS) {\it diverges} logarithmically and hence strongly enhances the effect of interactions. In recent years, several efforts have been devoted along this direction and possible chiral/helical SC and magnetic orders are predicted in various lattice structures.\cite{Gonzalez08,Martin08,McChesney10,Li12,Chubukov12,Gonzalez13,bSchaffer14,Yao15,Chen15,Meng15} 

According to a recent work by Yao and Yang,\cite{Yao15} one can separate VHS into two types by their position $\vec{K}$ in the first Brillouin zone (BZ): When $\vec{K}$ locate at time-reversal invariant momenta, VHS belong to type-I; otherwise, they belong to type-II. An immediate consequence is that, within the same crystal structure, the number of VHS of type-II is doubled, compared to that of type-I. This leads to different emergent symmetry breaking orders. For instance, in the hexagonal system like graphene, a chiral $d$-wave {\it singlet} SC is predicted\cite{Chubukov12}; however, in BC$_3$, a helical $p$-wave {\it triplet} SC is singled out.\cite{Chen15} So far, most of the studies assume spin SU(2) symmetry is preserved (or at least approximately) with each band doubly degenerate, while systems with strong spin-orbit coupling (SOC) such as usually seen on the surfaces of a 3D TI have not yet been discussed.

In this work, we study the physics around type-II VHS appearing in the topologically protected surface states of the Sb(111) thin films (or bulk), which commonly have a hexagonal surface BZ.\cite{Teo08,ARPES-Sb0,ARPES-Sb,ARPES-Sb2,Zhang12} We consider a $k\cdot p$ model for the surface states and perform patch renormalization group (RG) analysis to investigate the competing orders in the system near VHS. We show that the leading instability is a chiral $p+ip$-wave superconducting order under weak repulsive interactions, robust against the interaction strength and a range of material-dependent paramters. The origin of such pairing relies on the strong anisotropy of the electron-electron scattering after renormalization to provide an attractive channel, just in the same spirit of Kohn-Luttinger type mechanism.\cite{Kohn65,Gonzalez08} Our results therefore suggest that the Sb thin films could be another promising candidate for hosting Majorana zero modes, whcih may be utilized in performing topological quantum computation.\cite{Kitaev03,Nayak08} 

The paper is organized as follows. In Sec. II, we briefly discuss the model used to describe the surface states in the Sb(111) thin film and show the existence of the VHS. Sec. III turns to consider the low-energy effective theory of the system and sketch how we perform the RG analysis. After solving the RG equations and calculating the renormalized susceptibilities for various symmetry breaking orders, in Sec. IV, we present the phase diagram and discuss the leading order. Finally, Sec. V comments on experimental realization and the effect from the other possible form of interactions, and concludes with a summary of our results.

\begin{figure}
\includegraphics[scale=0.5]{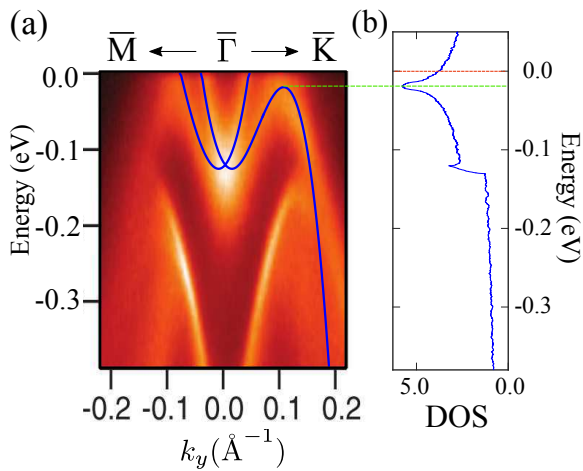}
\caption{(Color online) (a) The $k\cdot p$ model fits to the ARPES result in Ref.~\onlinecite{ARPES-Sb}. Note that the ARPES result does not include data along $\bar{\Gamma}\bar{M}$. (b) Density of states calculated from the $k\cdot p$ model, consistent with STM ($dI/dV$) results in Ref.~\onlinecite{STM-Sb}. The red line indicates the Fermi level given in ARPES result and the green one indicates the energy level of VHS.}
\label{fig: ARPESfit}
\end{figure}

\section{Model}
Bulk Sb is known as a topological semimetal with non-trivial surface states.\cite{Teo08} Making it a thin film, viewed as a stacking of (111) bilayers (BL), still preserves the topological property of the surface states so long as the number of BL is no less than five.\cite{Zhang12,Pan15,QPI-Sb} Thus, one can start with a bulk sample. Sb has the rhombohedral A7 structure, which consists of two interpenetrating, face-centered cubic lattices, displaced with each other along [111] direction. Similar to Bi$_2$Te$_3$, the crystal symmetry of the (111) surface is reduced to point group $C_{3v}$. Based on this fact, instead of developing a Liu-Allen-like tight-binding model,\cite{Liu-Allen} we simply adopt a $k\cdot p$ model, originally developed by Liang Fu for Bi$_2$Te$_3$,\cite{Fu09} for the surface electrons of Sb(111). It is essential to note that the form of such model is restricted by both $C_{3v}$ and TRS. Consequently,
%Thus, we start with a bulk Sb to construct a simple model for surface states. Sb has the rhombohedral A7 structure, which consists of two interpenetrating, face-centered cubic lattices, which are displaced with each other along [111] direction. On the (111) surface, the crystal structure is reduced to point group $C_{3v}$, which includes a $C_3$ rotation with respect to $z$-axis (\ie, [111] direction) and three reflection operations along $\bar{\Gamma}\bar{M}$ mirror lines. Instead of developing a Liu-Allen-like tight-binding model,\cite{Liu-Allen} we, for simplicity, adopt a $k\cdot p$ model expanding around $\bar{\Gamma}$ for the surface electrons of Sb(111).\cite{Fu09} This model is strictly fixed by $C_{3v}$ symmetry and time-reversal symmetry and is of the form
\be
H(\vec{k})=E_0(\vec{k})+v_{\vec{k}}(k_x \sigma_y-k_y\sigma_x)+\frac{\lambda}{2}(k_+^3+k_-^3)\sigma_z,
\ee
where $k_{\pm}=k_x\pm i k_y$, expanding from $\bar{\Gamma}$, $v_{\vec{k}}=v(1+\tilde{\alpha} k^2)$ denoting Fermi velocity with momentum-dependent corrections, and a natural pseudo-spin doublet at $\bar{\Gamma}$ corresponding to total angular momentum $J=L+S=\frac{1}{2}$ is chosen to be the basis. The first term, $E_0(\vec{k})=\frac{k^2}{2m^*}$, introduces particle-hole asymmetry, while the second and third terms correspond to SOCs in linear and cubic orders, respectively.

The (surface) band dispersion of the model can be easily solved as
\be
E_{\pm}(\vec{k})=E_0(\vec{k})\pm\sqrt{v_{\vec{k}}^2 k^2+\lambda^2 k^6 \cos^2(3\theta)},
\ee
where $\theta$ is the polar angle of $\vec{k}$ with respect to $k_x$ axis [defined in Fig. 2(a)]. These dispersion relations are then used to fit with the angle-resolved photoemission spectroscopy (ARPES) result\cite{ARPES-Sb}, as shown in Fig. 1(a), in order to obtain the optimized model parameters: $(2m^*)^{-1}=40$ eV$\cdot\text{\AA}^2$, $v=0.9$ eV$\cdot\text{\AA}$, $\tilde{\alpha}=137$ $\text{\AA}^2$, and $\lambda=210$ eV$\cdot\text{\AA}^3$ ($\hbar=c\equiv 1$ have been absorbed into parameters).  As an independent check, we find that in Fig. 1(b) our computed DOS of the model also fits well with scanning tunneling microscopy (STM) $dI/dV$ spectrum\cite{STM-Sb}, except for a possible Fermi energy shift due to sample or environment varience.

Upon hole doping, the surface Fermi surface (FS) of Sb(111) would undergo a Lifshitz transition, reflecting the fact that six 2D Van Hove singularities (saddle points) are present inside the surface BZ along high symmetry lines ($\bar{\Gamma}\bar{K}$ or $\bar{\Gamma}\bar{K}^\prime$), as shown in Fig. 2(a). Within our model for Sb(111), the VHS position in $k$-space can be estimated as $|\vec{K}_\alpha| \approx [3\sqrt{m^{*2}(\lambda^2+v^2\tilde{\alpha}^2)}]^{-1} $ with VHS indices $\alpha$=1 to 6.

\begin{figure}
\includegraphics[scale=0.29]{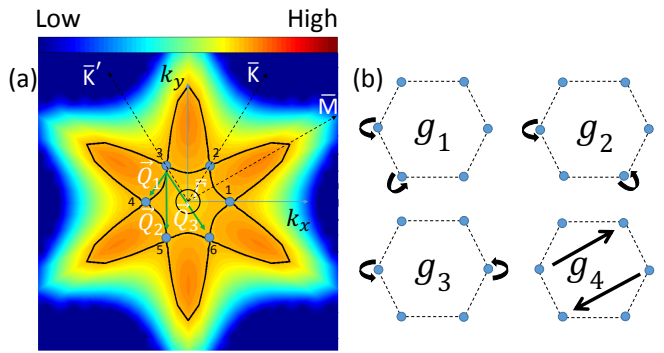}
\caption{(Color online) (a) The colored energy contour plot for the lower band in the $k\cdot p$ model.   There are six saddle points (blue dots) with hexagonal symmetry, locating along $\bar{\Gamma}\bar{K}$ and $\bar{\Gamma}\bar{K}^\prime$. Surface Fermi surface at the energy level of VHS is shown in bold black curve. Three inequivalent nesting vectors $\vec{Q}_j$ are also indicated by green arrows. Note that the `low' and `high' labels of the color bar are relative to the energy level of VHS. (b) A schematic plot for the four distinct interactions in our system within patch (blue dots) approximation.}
\label{fig:FS}
\end{figure}

\section{Patch renormalization group analysis}
A {\bf salient} feature for a 2D VHS is that its DOS is logarithmically {\it divergent}, suggesting Fermi liquid instabilities in the presence of even `weak' electron-electron interactions whenever the Fermi level is around the saddle points. Focusing only on the low-energy physics in the weak interaction regime, we therefore legitimately take the ``patch approximation''\cite{Chubukov12,Yao15,Chen15} and neglect electrons far away from the saddle points. 

\subsection{Effective theory}
Taking into account the patches around six saddle points, the low-energy physics can be described by the following effective action:
\ba
S &=&\int d\tau d^2 x\sum_{\alpha=1}^{6}\{ \psi^\dagger_\alpha\left[-\partial_\tau-\epsilon_{\alpha}(-i\partial_x,-i\partial_y)+\mu\right]\psi_\alpha \nonumber \\ 
&-& \frac{g_1}{2}\psi^\dagger_\alpha\psi^\dagger_{\alpha^\prime}\psi_{\alpha^\prime}\psi_{\alpha} -\frac{g_2}{2}\psi^\dagger_\alpha\psi^\dagger_{\alpha^{\prime\prime}}\psi_{\alpha^{\prime\prime}}\psi_{\alpha}-\frac{g_3}{2}\psi^\dagger_\alpha\psi^\dagger_{\bar{\alpha}}\psi_{\bar{\alpha}}\psi_{\alpha} \nonumber \\
&-& \frac{g_4}{2}[\psi^\dagger_\alpha\psi^\dagger_{\bar{\alpha}}\psi_{\alpha+1}\psi_{\bar{\alpha}+1}+h.c.]\}, 
\label{eq:effective}
\ea
where $\psi^\dagger_{\alpha}$ are creation operators of electrons with patch (VHS) indices $\alpha=1,\cdots$, 6 [see Fig. 2(a)]. Note that due to strong SOC, each band is non-degenerate and thus electrons here are effectively considered {\it spinless}. The indices $\alpha^\prime$, $\alpha^{\prime\prime}$, and $\bar{\alpha}$ label nearest-neighbor (NN), next nearest-neighbor (NNN), and third neighbor (NNNN) patchs of $\alpha$, respectively. For a given patch, \eg, $\alpha=1$, the energy dispersion around the saddle point is $\epsilon_1(\vec{q}) =-\frac{q_x^2}{2 m_x}+\frac{q_y^2}{2 m_y}$, where $\vec{q}=\vec{k}-\vec{K}_1$. The DOS per patch is then easily calculated as $N(\omega)\approx 2N_0\ln\frac{\Lambda}{\omega}$, where $N_0=\frac{\sqrt{m_x m_y}}{4\pi^2}$, $\Lambda$ is order of surface band width, and $\omega$ is the energy away from VHS. The dispersions around the other saddle points can be obtained by $C_6$ operations on $\vec{q}$; inequivalent saddle points are connected by three types of nesting vectors: $\vec{Q}_{j}=\vec{K}_{j+1}-\vec{K}_1$ for $j=1,2,3$ [see Fig. 2(a) and $\vec{Q}_0\equiv 0$]. In our system, we have $m_x\approx 0.0137$ $\text{\AA}^{-2}\cdot\text{eV}^{-1}$ and $m_y\approx 0.0071$ $\text{\AA}^{-2}\cdot\text{eV}^{-1}$. Here we take the chemical potential $\mu=0$, which describes a system doped exactly to the saddle points.

In the effective theory, the short-range interaction is assumed, which may be justified by the metallic screening effect due to the states near the Fermi surface. There are four types of interactions, denoting their coupling strength from $g_1$ to $g_4$ [see Fig. 2(b)], which are constrained by the momentum conservation. The first three types represent NN, NNN, and NNNN density-density interactions and thus the bare values of $g_1$, $g_2$, and $g_3$ are generically positive. The fourth type represents the pair hopping process. All of these interactions are {\it marginal} at tree level in 2D and would acquire logarithmic corrections (divergences) in perturbation theory. Therefore, below, we will employ the RG technique\cite{Shankar94} to deal with this situation and to determine which kind of symmetry breaking order might occur as the temperature decreases.

\subsection{RG equations}
We perform RG analysis up to one-loop level via integrating out the high-energy degrees of freedom gradually from the energy cutoff $\Lambda$ to study how interactions flow. Practically, the essential first step is to study various (non-intereacting) bare susceptibilities in both particle-hole and particle-particle channels. Each susceptibility is a kind of measure of the nesting property between patches connected by $\vec{Q}_j$ at a given $\omega$. The physical consequence of the comparison with the nesting property would reflect on the enhancement of anisotropy among different electron-electron interactions, $g_i$. Note that only the susceptibilities in the Cooper channel can have log-square behavior:
\ba
\chi_{\vec{Q}_0}^{pp}(\omega) &\approx & N_0\ln^2\frac{\Lambda}{\omega},\quad
\chi_{\vec{Q}_0}^{ph}(\omega)  \approx  2N_0\ln\frac{\Lambda}{\omega}, \\
\chi_{\vec{Q}_1}^{pp}(\omega) &\approx & 2N_0\bar{a}\ln\frac{\Lambda}{\omega},\quad
\chi_{\vec{Q}_1}^{ph}(\omega) \approx 2N_0 a\ln\frac{\Lambda}{\omega}, \\
\chi_{\vec{Q}_2}^{pp}(\omega) &\approx & 2N_0\bar{a}\ln\frac{\Lambda}{\omega},\quad 
\chi_{\vec{Q}_2}^{ph}(\omega) \approx 2N_0 a\ln\frac{\Lambda}{\omega}, \\
\chi_{\vec{Q}_3}^{pp}(\omega) &\approx & N_0 a_3\ln^2\frac{\Lambda}{\omega},\quad
\chi_{\vec{Q}_3}^{ph}(\omega) \approx 2N_0 a_3\ln\frac{\Lambda}{\omega},
\ea
where $N_0$ and $\vec{Q}_j$ are defined in previous subsection. $a$ and $\bar{a}$ are functions of mass ratio $\kappa=\frac{m_y}{m_x}$, while $0<a_3\leqslant 1$ dependes on the detailed dispersion mismatch. The subscript of $\chi^{pp}$ indicates the total momentum of a particle-particle pair, but that of $\chi^{ph}$ represents the momentum transfer of a particle-hole bubble. Note that $\chi_{\vec{Q}_1}^{pp(ph)}=\chi_{\vec{Q}_2}^{pp(ph)}$ are due to hexagonal symmetry of our system. For Sb(111) in the effective theory, a numerical estimation gives $\kappa\approx 0.52$, $a\approx 2$, $\bar{a}\approx 1$, and $a_3\approx 1$. 
%(may discuss sensitivity of these nesting parameters to the final result...)

\begin{figure}
\includegraphics[scale=0.32]{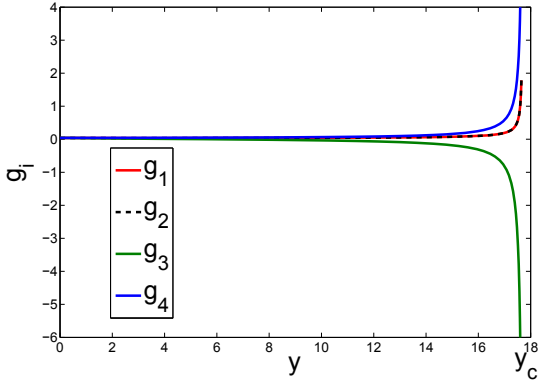}
\caption{(Color online) Flow of couplings with RG scale $y$, starting from repulsive bare values, $g_i=0.04$, and model parameters, $a=2$ and $\bar{a}=a_3=1$. Note that the coupling $g_3$ changes sign eventually, leading to a superconducting instability at the scale $y_c$.}
\label{fig:g-running}
\end{figure}

With logarithmic accuracy, using $y=\ln^2(\frac{\Lambda}{\omega})$ as the RG flow time, the derived RG equations are given as follows:
\ba
\frac{d g_1}{dy} &=& -d_2^{pp}g_{1}^2 + d_1^{ph}(g_1^2+g_4^2) - 2d_0^{ph}(g_1 g_2+g_2 g_3),\nonumber \\ \frac{d g_2}{dy} &=& -d_1^{pp}g_2^2 + d_2^{ph}(g_2^2+g_4^2) - d_0^{ph}(2g_1 g_3+g_1^2+g_2^2),\nonumber \\
\frac{d g_3}{dy} &=& -(g_3^2+2g_4^2) + d_3^{ph}g_3^2 - 4d_0^{ph}g_1 g_2,\nonumber \\
\frac{d g_4}{dy} &=& -(g_4^2+2g_3 g_4) + 2d_1^{ph}g_1 g_4 + 2d_2^{ph}g_2 g_4, 
\label{RGeqs}
\ea
where each $g_i$ represents a dimensionless coupling strength by introducing $g_i\rightarrow N_0 g_i$. Here we define the (relative) ``nesting parameters'' $d_\mu^{ph(pp)}=\partial\chi^{ph(pp)}_{\vec{Q}_\mu}/\partial\chi^{pp}_{\vec{Q}_0}$ with $\mu=0,1,2,3$.\cite{Chubukov12,Yao15,Chen15} They are decreasing functions of $y$ and have the following asymptotic behavior: $d_\mu^{ph(pp)}\rightarrow 1$ as $y\rightarrow 0$; $d_1^{pp}\rightarrow \frac{\bar{a}}{\sqrt{y}}$, $d_2^{pp}\rightarrow \frac{\bar{a}}{\sqrt{y}}$, $d_3^{pp}\rightarrow a_3$, $d_0^{ph}\rightarrow \frac{1}{\sqrt{y}}$, $d_1^{ph}\rightarrow \frac{a}{\sqrt{y}}$, $d_2^{ph}\rightarrow \frac{a}{\sqrt{y}}$, $d_3^{ph}\rightarrow \frac{a_3}{\sqrt{y}}$ as $y\rightarrow \infty$.

We integrate our RG equations in Eq. (\ref{RGeqs}) and model $d_\mu^{ph(pp)}$ as $d_1^{pp}(y)\approx \frac{\bar{a}}{\sqrt{\bar{a}^2+y}}$, $d_2^{pp}(y)\approx \frac{\bar{a}}{\sqrt{\bar{a}^2+y}}$, $d_3^{pp}(y)\approx \frac{1+a_3 y}{1+y}$, $d_0^{ph}(y)\approx \frac{1}{\sqrt{1+y}}$, $d_1^{ph}(y)\approx \frac{a}{\sqrt{a^2+y}}$, $d_2^{ph}(y)\approx \frac{a}{\sqrt{a^2+y}}$, and $d_3^{ph}(y)\approx \frac{a_3}{\sqrt{a_3^2+y}}$ to smoothly interpolate between the limiting aymptotic behaviors. As illustrated in Fig. 3, we observe that $g_i(y)$ typically flow to infinity as $y\rightarrow y_c$ and thus can be scaled as 
\be
g_i(y)\sim\frac{G_i}{y_c-y}, \label{eq:gasymp}
\ee
where $G_i$ is a constant, when close to $y_c$. Such divergences indicate that the system evolves to strong coupling regime and certain instability would occur at the energy scale 
\be
\omega_c\sim\Lambda e^{-\sqrt{y_c}}, \label{eq:tc}
\ee
as we will discuss next.

%Solving the $\beta$-function numerically, setting all initial values ${{g}_{i}}=0.02$ and $a=\bar{a}=a_3=1$, This boundary condition is selected for illustration. The larger the initial ${{g}_{i}}$  setting results in a higher transition temperature this many provide us some insight of high temperature superconductors and the result is show in Fig 2. we can tell that ${{g}_{3}}$ begin with positive value but diverge into negative at the point ${{y}_{c}}\approx 60$. This means that by lowering the temperature, ${{g}_{3}}$ channel become effectively attractive. Those coupling sign changing behavior will cause Fermi surface unstable then lead to the system symmetry braking into some lower energy states spontaneously. Meanwhile the value of $g_1$ and $g_2$ are very close but not equal so we can only see three lines in Fig 2.

\section{Competing orders and phase diagram}
Following the same strategy used in Refs.~\onlinecite{Chubukov12,Yao15,Chen15,Vafek14}, one may map out the qualitative phase diagram of the system by evaluating the renormalized susceptibilities, which diverge like $(y_c-y)^\alpha$, for various types of symmetry breaking orders. When lowering the temperature, the actual order would occur at the phase transition with the most negative power exponent $\alpha$. 

\subsection{Renormalized susceptibilities}
Among all the orders we have investigated in the system, we find that the superconducting instability is the most dominant one as long as the bare electron-electron interactions are repulsive. Therefore, we sketch our analysis on superconducting instability (with Cooper pair momentum zero) as an illustrative example.

We first add test infinitesimal vertices in the particle-particle channel into Eq. (\ref{eq:effective}),
\be
\delta\mathcal{L}_{0-SC}=\sum_{\alpha=1}^{6}[\Delta_{\alpha\bar{\alpha}}\psi_{\alpha}^{\dagger}\psi_{\bar{\alpha}}^{\dagger} + h.c.].
\ee
The renormalization of the test vertices is then governed by the following matrix equation,
\be
\frac{d}{dy}\left(\begin{matrix}
\Delta_{14}  \\
\Delta_{41}  \\
\Delta_{25}  \\
\Delta_{52}  \\
\Delta_{36}  \\
\Delta_{63}  \\
\end{matrix} \right)=2\left(\begin{matrix}
   -g_3 & g_3 & g_4 & -g_4 & -g_4 & g_4  \\
   g_3 & -g_3 & -g_4 & g_4 & g_4 & -g_4  \\
   g_4 & -g_4 & -g_3 & g_3 & g_4 & -g_4  \\
   -g_4 & g_4 & g_3 & -g_3 & -g_4 & g_4  \\
   -g_4 & g_4 & g_4 & -g_4 & -g_3 & g_3  \\
   g_4 & -g_4 & -g_4 & g_4 & g_3 & -g_3  \\
\end{matrix}\right)\left(\begin{matrix}
\Delta_{14}  \\
\Delta_{41}  \\
\Delta_{25}  \\
\Delta_{52}  \\
\Delta_{36}  \\
\Delta_{63}  \\
\end{matrix}\right), \label{eq:rgforchi}
\ee
which can be diagonalized to obtain the eigenmodes. Ignoring three unphysical eigenmodes with zero eigenvalue, there are three eigenmodes $w_j$ (eigenvalues $\varepsilon_j$),
\ba
&w_f&=\frac{\Delta_f}{\sqrt{6}}\left(\begin{matrix}
	                                  1  \\
	                                 -1  \\
	                                 -1  \\
	                                  1  \\
	                                  1  \\
	                                 -1  \\
\end{matrix} \right),
w_{p_x}=\frac{\Delta_{p_x}}{2\sqrt{3}}\left(\begin{matrix}
	                                        2  \\
	                                       -2  \\
	                                        1  \\
	                                       -1  \\
	                                       -1  \\
	                                        1  \\
\end{matrix} \right),
w_{p_y}=\frac{\Delta_{p_y}}{2}\left(\begin{matrix}
	                                0  \\
	                                0  \\
	                                1  \\
	                               -1  \\
	                                1  \\
	                               -1  \\
\end{matrix} \right), \nonumber \\
&\varepsilon_f& = -(4g_4+2g_3),\quad\varepsilon_{p_x}=\varepsilon_{p_y}=2(g_4-g_3),\label{eq:eigenmode}
%&w_f& = \frac{\Delta_f}{\sqrt{6}}(1\,-1\,-1\,1\,1\,-1)^t, \varepsilon_f=-(4g_4+2g_3),\label{eq:wf}\\
%&w_{p_x}& =\frac{\Delta_{p_x}}{2\sqrt{3}}(2\,-2\,1\,-1\,-1\,1)^t, %\varepsilon_{p_x}=2(g_4-g_3),\label{eq:wpx}\\
%&w_{p_y}& = \frac{\Delta_{p_y}}{2}(0\,0\,1\,-1\,1\,-1)^t, \varepsilon_{p_y}=\varepsilon_{p_x}\equiv %\lambda,\label{eq:wpy}
\ea
corresponding to $f$-wave superconductivity in $A_2$ irreducible representation of $C_{3v}$ and degenerate $p_x/p_y$-wave superconductivity in $E$ irreducible representation, respectively. Within the eigenmode basis, each order parameter $\Delta_j$ ($j=f,p_x,p_y$) now obeys $\frac{d\Delta_j}{dy}=2\varepsilon_j\Delta_j$. Inserting the scaling form of $g_i$ in Eq. (\ref{eq:gasymp}) and the susceptibility of such order $\chi_j(y)=\Delta_j(y)/\Delta_j(0)\sim(y_c-y)^{\alpha_j}$ to it, one obtains $\alpha_{f-SC}=4(G_3+2G_4)$ and $\alpha_{p_x-SC}=\alpha_{p_y-SC}=4(G_3-G_4)$.

\begin{figure*}[th]
\begin{center}
\subfigure[]{
%\label{fig:suscep:a}
\includegraphics[scale=0.35]{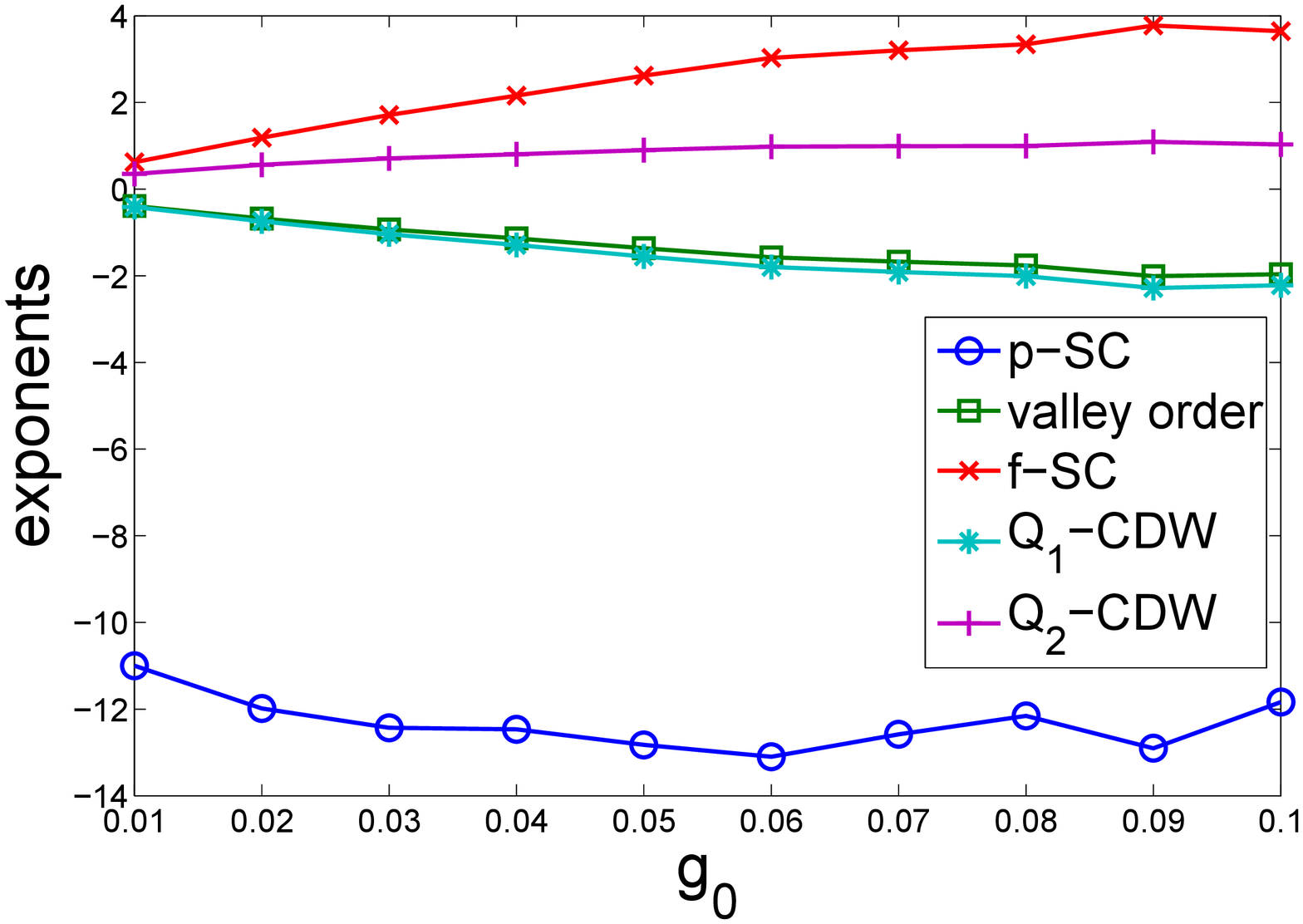}}
%\hspace{1in} 
\subfigure[]{
%\label{fig:suscep:b}
\includegraphics[scale=0.35]{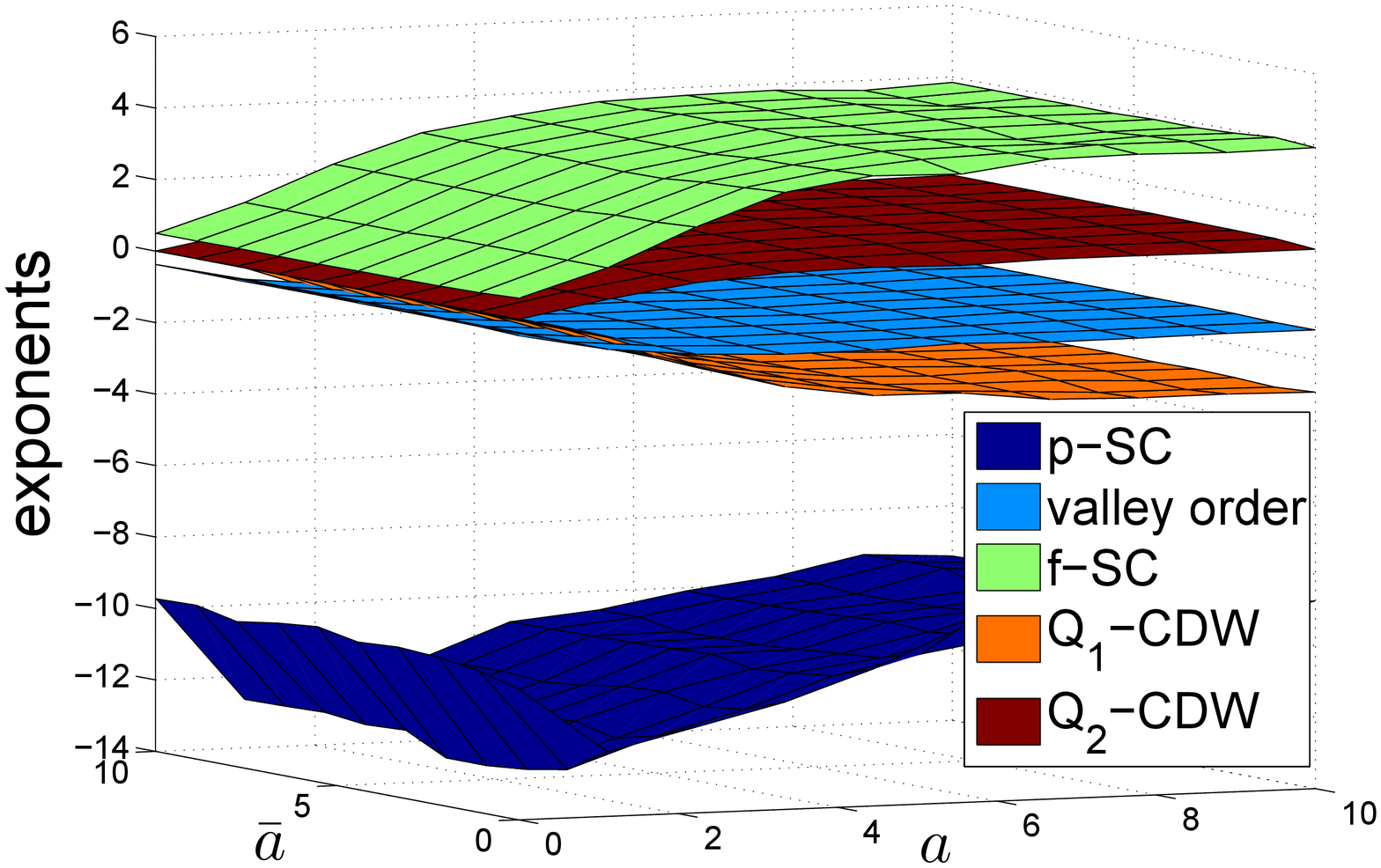}}
\caption{(Color online) The susceptibility exponents of various types of symmetry breaking orders as a function of (a) $g_i=g_0$ and (b) $a$ and $\bar{a}$. In (a), the chosen model parameters are $a=2$ and $\bar{a}=a_3=1$; In (b), the bare value of $g_0$ is set to be 0.04. The most negative power exponent indicates the leading order of the system. Clearly, the dominant $p$-wave SC is shown to be robust against either the change of $g_0$ or the change of the model (nesting) parameters.}
\label{fig:suscep}
\end{center}
\end{figure*}

In fact, applying the same approach sketched above, one can obtain the other finite momentum (either $\vec{Q}_1$ or $\vec{Q}_2$) pairing instabilities, associated with the exponents $\alpha_{FFLO}^{(1)}=4d^{pp}_1 G_2$ and $\alpha_{FFLO}^{(2)}=4d^{pp}_2 G_1$, respectively. Moreover, there are also competing orders in particle-hole channel such as (i) valley orders ({\it i.e.} density imbalance among different patches without breaking crystal translation symmetry) with $\alpha_{v1}=2d^{ph}_0(-2G_1+2G_2-G_3)$, $\alpha_{v2}=2d^{ph}_0(G_1-G_2-G_3)$, and $\alpha_{v3}=2d^{ph}_0(-G_1-G_2+G_3)$; (ii) charge density wave (CDW) orders with $\alpha_{Q_1-CDW}^{(1)}=-2d^{ph}_1(G_1+G_4)$, $\alpha_{Q_1-CDW}^{(2)}=-2d^{ph}_1(G_1-G_4)$, $\alpha_{Q_2-CDW}^{(1)}=-2d^{ph}_2(G_2-G_4)$, $\alpha_{Q_2-CDW}^{(2)}=-2d^{ph}_2(G_2+G_4)$, and $\alpha_{Q_3-CDW}^{(1)}=-2d^{ph}_3 G_3$. 

Given the values of $a$, $\bar{a}$, and $a_3$, already mentioned in Sec. IIIB, we obtain various leading susceptibility exponents as a function of $g_i=g_0>0$ in Fig. \ref{fig:suscep}(a). Near $y_c$, $g_3$ quickly goes to -$\infty$, while $g_1$, $g_2$, and $g_4$ go to $\infty$ [see Fig. \ref{fig:g-running}]. As a consequence, the most prominent leading instability is the doubly degenerate $p$-wave pairing regardless of the strength of $g_0$. As we will show later, it turns out to be the chiral $p+ip$ pairing. In Fig. \ref{fig:suscep}(b), we further investigate the sensitivity of this result under changing the material-dependent, nesting properties $a$ and $\bar{a}$ with $a_3=1$. We observe that $p$-wave pairing is still robust. Note that our result is in contrast to the systems with hexagonal type-I VHS where either chiral $d$-wave pairing or certain magnetic ordering is favored.\cite{Chubukov12,Li12} Instead, our result is relatively closer to that obtained in the systems with hexagonal type-II VHS and unbroken spin SU(2) symmetry,\cite{Chen15} although here all magnetic orders are gone and chiral, rather than helical, superconductivity is performed. All of the distinctions are mainly due to the differences in the total number of VHS and the presence of large SOC. 

One final remark is worth mentioning here. When the bare $g_4$ is set to be zero, it flows to zero eventually, resulting in the degeneracy between $p$-wave and $f$-wave pairings [see also Fig. 5]. The presence of bare positive (negative) value of $g_4$ would pick up $p$-wave ($f$-wave) pairing. Since $g_4$ describes hopping processes of a time-reversal invarinat Cooper pair to NN or NNN saddle points, combining the fact that from spin texture around each saddle point NN (NNN) saddle points have opposite (same) out-of-the-plane spin polarizations, positiveness of $g_4$ is more likely achieved. This is consistent with Ref.~\onlinecite{Chen15}, in which the comparison of the sub-logarithic behavior in scattering channels with $\vec{Q}_1$ and $\vec{Q}_2$ is essential to determine the dominant $p$-wave or $f$-wave pairing.

\subsection{Competition of $p$-wave orders below $T_c$}
For Sb (111) thin films, we have found that in many cases $p$-wave pairing is favored, as discussed in the previous subsection. However, $p_x$ and $p_y$ pairings are degenerate within $E$ irreducible representation of the crystal symmetry. In other words, any linear combination of them may also be an allowed solution with the same pairing susceptibility. From energetics, it is generally expected that the most promising combination is the one which gaps out the whole Fermi surface and gains more condensation energy.\cite{Cheng10} To determine the solution explicitly, one can analyze the Ginzburg-Landau free energy of the system when both $p_x$ and $p_y$ superconducting (SC) order parameters are present.

We start with the partition function of our system in the path integral formalism: $\mathcal{Z}=\int\mathcal{D}[\bar{\Psi},\Psi]\mathcal{D}[\bar{\Delta},\Delta]\text{exp}(\int\mathcal{L}_{MF}[\bar{\Psi},\Psi,\bar{\Delta},\Delta])$, where $\mathcal{L}_{MF}$ describes the Lagrangian density after mean-field decoupling of the original interactions to introduce degenerate SC order parameters in Nambu space
\ba
\mathcal{L}_{MF} &=& \bar{\Psi}\hat{M}\Psi+\frac{|\Delta_x|^2+|\Delta_y|^2}{\lambda}, \nonumber \\
\hat{M} &=& \left(\begin{matrix}
	G_p^{-1} & \Delta  \\
	\bar{\Delta} & G_h^{-1} \\
\end{matrix}\right),
\ea
with $\bar{\Psi}=(\bar{\psi}_1\,\bar{\psi}_2\,\bar{\psi}_3\,\psi_4\,\psi_5\,\psi_6)$ and the matrix, $\Delta=\Delta_x\hat{P_x}+\Delta_y\hat{P_y}$, in which the complex order parameters are defined as [via Eq. (\ref{eq:eigenmode})]
\ba
\Delta_x &=&\frac{\lambda}{\sqrt{6}}\left\langle 2\psi_4\psi_1+\psi_5\psi_2-\psi_6\psi_3\right\rangle, \nonumber \\
\Delta_y &=&\frac{\lambda}{\sqrt{2}}\left\langle \psi_5\psi_2+\psi_6\psi_3\right\rangle,
\ea 
and diagonal matrices $\hat{P_x}=\frac{1}{\sqrt{6}}\text{diag}(2,1,-1)$; $\hat{P_y}=\frac{1}{\sqrt{2}}\text{diag}(0,1,1)$. For each patch, the particle and hole Green's functions are given by $G_{p(h)}=[i\omega_n\mp(\epsilon-\mu)]^{-1}$.

By integrating out fermionic degrees of freedom, we get the effective action in terms of the SC order parameters up to quartic order:
\ba
\mathcal{L}_{GL} &=& -\text{Tr}\ln\hat{M}+\frac{1}{\lambda}(|\Delta_x|^2+|\Delta_y|^2) \nonumber \\
&=& \text{Tr}[G_p\Delta G_h\bar{\Delta}]+\frac{1}{2}\text{Tr}[G_p\Delta G_h\bar{\Delta}G_p\Delta G_h\bar{\Delta}] \nonumber \\
&+& \frac{1}{\lambda}(|\Delta_x|^2+|\Delta_y|^2)+\cdots \nonumber \\
&=& r(|\Delta_x|^2+|\Delta_y|^2)+u_1(|\Delta_x|^2+|\Delta_y|^2)^2 \nonumber \\
&+& 2u_2|\Delta_x|^2 |\Delta_y|^2 + u_3|\Delta_x\Delta_y^*+\Delta_x^*\Delta_y|^2+\cdots,
\ea
where the trace above includes the integration over momentum and the coefficients $r=\text{Tr}[G_pG_h]+\lambda^{-1}$, $u_1=\frac{1}{4}\text{Tr}[G_pG_hG_pG_h]>0$, $u_2=-\frac{1}{6}\text{Tr}[G_pG_hG_pG_h]<0$, $u_3=\frac{1}{12}\text{Tr}[G_pG_hG_pG_h]>0$. The sign of $u_2$ indicates that $p_x$-wave and $p_y$-wave can coexist, and the signs of $u_1$ and $u_3$ enforce us to conclude $\Delta_y=\pm i\Delta_x$. This confirms our expectation of the presence of $p+ip$ superconductivity.

\begin{figure}[t]
\includegraphics[scale=0.26]{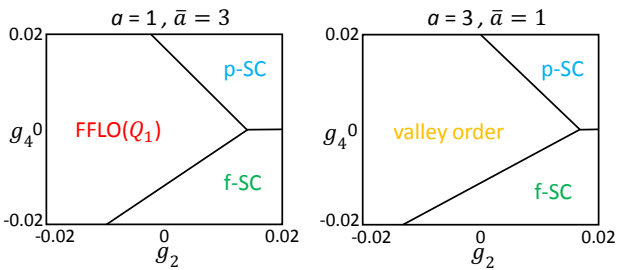}
\caption{(Color online) The phase diagrams as a function of $g_4$ and $g_2$ under the settings (a) $a=3>\bar{a}=1$ and (b) $a=1<\bar{a}=3$, with bare values $g_1=g_3=0.02$ in both cases.}
\label{fig:phasediagram}
\end{figure}

\section{Discussion and conclusion}
The feasibility of our proposal for VHS physics realizing in Sb(111) thin films may rely on the following two conditions: (i) the ability to tune the chemical potential to approach the VHS and (ii) suppressing the possible scattering from the bulk states. The former concern usually can be overcome by varying the gate voltage. As indicated in Fig. \ref{fig: ARPESfit}(a), the energy shift of Fermi level could be less than 100 meV, in sharp contrast to around 2.5 eV from the charge neutrality point\cite{Chubukov12} in graphene and around 1 eV in BC$_3$.\cite{Chen15} The latter issue is more severe. However, on the one hand, we notice that for the thick thin films at the level of VHS the bulk (quantum well) states usually appear along $\bar{\Gamma}\bar{M}$ with crystal momentum much larger than $K_\alpha$. With the Coulomb interaction $V(q)\sim\frac{1}{q}$ in mind, one may argue that any scattering between states around a saddle point and the bulk states, requiring larger momentum transfer $q$, is relatively weaker than that between saddle points. On the other hand, a better way to resolving this difficulty is to control the thickness of the thin film. From first-principles calculations,\cite{Zhang12} people have predicted that the Sb(111) thin film exhibits 3D TI phase (no bulk states around VHS) when its thickness ranging from around 3 nm to 7 nm (1 BL$\sim 3.75 \AA$). As a byproduct, the natural Fermi level also shifts from slightly below VHS (3 nm) to slightly above VHS (7 nm). In other words, selecting an appropriate thickness for the thin film may solve both issues at the same time. In addition, the presence of an insulating or semiconducting substrate in growing the thin film also brings an advantage that the VHS from the upper and lower surfaces could be separated in energy and may safely neglect the effect from quantum tunneling due to small DOS from the other side.

In the perspective of applications, it is quite useful to give a quick $T_c$ estimation for the superconductivity in the thin films. Using Eq. (\ref{eq:tc}) and the BCS relation between SC gap and $T_c$,\cite{Gonzalez08} the transition temperature can be roughly estimated as $k_B T_c\sim\frac{2}{1.76}\Lambda e^{-\sqrt{y_c}}$. Given a reasonable Coulomb repulsion $U$ up to few eV for relevant $p$-orbitals according to Refs.~\onlinecite{Chen15,Craco15}, it would correspond to $g_i=U N_0\approx\mathcal{O}(10^{-2})$. Taking $g_i=0.04$, for instance, with the surface band width around 0.2 eV$\sim\Lambda$,\cite{Zhang12} our numerical calculation yields $y_c\approx 18$ (see Fig.~3), giving $T_c\sim 30$K. This makes the possible application for quantum computing practical under current experimental technique.

One more remark on the form of weak interactions may deserve mentioning here. Although for the repulsive Hubbard-like interactions the bare values of $g_i$ are all positive, one could also imagine a more complicated form of interactions, for instance, due to certain spin/charge fluctuations or screening such that the interaction strength is oscillating ({\it e.g.}, $g_1,g_3>0$ while $g_2<0$). As a consequence, in Fig. \ref{fig:phasediagram}, we notice that some other broken symmetry order, rather than $p+ip$ superconductivity, can become the leading instability. Depending on better nesting property either in the particle-particle ($a<\bar{a}$) or particle-hole channel ($a>\bar{a}$), the FFLO pairing\cite{Fulde64,Larkin64} with finite momentum $\vec{Q}_1$ or valley imbalance charge order ($\alpha_{v1}$) emerges as the dominant one eventually.

%3) conclusion
In summary, we have performed RG analysis for a hexagonal system with large SOC, such as the Sb(111) thin film, close to the type-II VHS. We find that such system has the leading instability to exhibiting $p+ip$ superconducting order from purely repulsive interactions and, in particular, the emergent SC order is quite robust against material-dependent, nesting-related parameters and the interaction strength. Moreover, such SC order is also known to host chiral edge modes and Majorana zero modes within the magnetic half-vortices. Thus, our present work adds a new potential use of the Sb (111) thin films in the context of topological quantum computation, besides the usual proposals for electronics and spintronics based mainly on their topological surface states.\cite{Hasan10,Qi11,Ando13}

\begin{acknowledgments}
We thank C.-Y. Huang, S.-K. Jian, and H. Yao for useful discussions and the collaboration on a related topic. J.Q.H. and D.X.Y. acknowledge the support from NBRPC-2012CB821400, NSFC-11574404, NSFC-11275279, Natural Science Foundation of Guangdong Province (China)-2015A030313176, NSFC-Guangdong Joint Fund and National Supercomputer Center in Guangzhou, Shanhai Forum, and Fundamental Research Funds for the Central Universities of China. The work at National University of Singapore (Singapore) is supported by the National Research Foundation, Prime Minister’s Office, Singapore, under its NRF fellowship (NRF Award No. NRF-NRFF2013-03).
\end{acknowledgments}

\end{document}